\documentclass[aps,prl,twocolumn,showpacs,groupedaddress]{revtex4-1}

\usepackage{graphicx}
\usepackage{bm}

\begin{document}


\title{Full Counting Statistics for Orbital-Degenerate Impurity Anderson Model \\
with Hund's Rule Exchange Coupling}


\author{Rui Sakano$^1$, Yunori Nishikawa$^2$, Akira Oguri$^2$, Alex C. Hewson$^3$, and Seigo Tarucha$^1$}
\affiliation{$^1$Department of Applied Physics, University of Tokyo, Bunkyo, Tokyo, Japan \\
$^2$Department of Physics, Osaka City University, Sumiyoshi, Osaka, Japan \\
$^3$Department of Mathematics, Imperial College, London SW7 2AZ, United Kingdom}


\date{\today}

\begin{abstract}
We study nonequilibrium current fluctuations 
through a quantum dot, which includes 
a ferromagnetic Hund's rule coupling $J$,
in the low-energy Fermi liquid regime
using the renormalized perturbation theory.
The resulting cumulant for the current distribution
in the particle-hole symmetric case
shows that
spin-triplet and spin-singlet pairs of quasiparticles
are formed in the current
due to the Hund's rule coupling and these pairs enhance the current fluctuations.
In the fully screened higher-spin Kondo limit,
the Fano factor takes a value $F_b = (9M+6)/ (5M+4)$
determined by the orbital degeneracy $M$.
We also investigate the crossover between the small
and large $J$ limits
in the two-orbital case $M=2$,
using the numerical renormalization group approach.
\end{abstract}


\maketitle

The experimental study of  electron transport in  mesoscopic devices,
subject to applied bias voltages, has provided a new way of investigating
 interelectron many-body effects
 \cite{PhysRevLett.100.246601,PhysRevLett.67.3720,PhysRevB.64.153305}. 
For example, the
theoretical prediction that the
Kondo correlation on the nonequilibrium currents
through quantum dots leads to an enhancement of the shot noise in these systems
\cite{PhysRevLett.97.086601,PhysRevB.73.233310,PhysRevB.80.155322,JPSJ.79.044714} has stimulated subsequent experiments
\cite{PhysRevB.77.241303,NaturePhys5.208,PhysRevLett.106.176601}.

A deeper understanding of nonequilibrium transport
in these systems can be obtained from the calculation of
 the current distribution function,
the higher order cumulants
beyond averaged current, and  the noise power.
These can be calculated from 
the cumulant generating function (CGF)
for nonequilibrium transport. This quantity, however, is difficult  to calculate
when  many-body effects have to be taken into account.
A recent development has been the calculation of the current probability distribution
for quantum dot in the Kondo regime 
\cite{PhysRevLett.97.016602,PhysRevB.76.241307,PhysRevB.83.241301} 
using the general formulation of the full counting statistics (FCS)
\cite{PROP:PROP200610305,RevModPhys.81.1665}.
It was shown that
quasiparticle singlet pairs
carry a charge $2e$ 
in the backscattering current due to the Kondo correlation and that these 
enhance the nonequilibrium current fluctuations.

In this Letter, we investigate the FCS for
an orbital-degenerate impurity Anderson model
as a prototype model to examine the effects of both
the Hund's rule and Kondo correlations and, hence, to deduce
the time-averaged current and shot noise. These 
have been experimentally investigated
in vertical dots, carbon nanotubes,
and double dots
\cite{PhysRevLett.93.017205,NaturePhys5.208,PhysRevLett.101.186804}.
Using the renormalized perturbation theory (RPT)
\cite{PhysRevLett.70.4007},
we calculate
the zero-temperature CGF up to third order in
applied bias voltage
for arbitrary strength of
the interactions at the dot site 
in the low-energy Fermi-liquid regime.
We show that
the Hund's rule correlation gives rise to
spin-triplet pairs and spin-singlet pairs of quasiparticles
carrying charge $2e$ in the nonequilibrium current,
which characterize the shot noise and higher order cumulants 
at low energies.

{\it Model}.---
Let us consider transport through a correlated dot
described by an orbital-degenerate impurity Anderson model
${\cal H}_{A}^{} = {\cal H}_0^{} +{\cal H}_{T}^{}+{\cal H}_{I}^{}$ with
\begin{eqnarray}
{\cal H}_{0}^{} &=& \sum_{k\alpha m\sigma} \varepsilon_{k\alpha}^{} c_{k \alpha m\sigma}^{\dagger} c_{k\alpha m\sigma}^{}
 + \sum_{m\sigma}  \epsilon_{d}^{} \, n_{dm\sigma}^{} , \\
{\cal H}_{T}^{} &=& \sum_{k\alpha m\sigma} \left( v_{\alpha}^{} d_{m\sigma}^{\dagger} c_{k \alpha m\sigma}^{} + \mbox{H.c.} \right) , \\
{\cal H}_{I}^{} &=& U \sum_{m} n_{dm\uparrow}^{} n_{dm\downarrow}^{} + W \sum_{m>m',\sigma\sigma'} n_{dm\sigma}^{} n_{dm'\sigma'}^{} \nonumber \\
&& \quad + 2J \sum_{m>m'} {\bm S}_{dm}^{} \cdot {\bm S}_{dm'}^{}.
\end{eqnarray}
Here, $d_{m\sigma}^{}$ annihilates an electron
in the dot level $\epsilon_{d}^{}$ with
orbital $m=1,2,\cdots, M$ and spin $\sigma$,
and $c_{k\alpha m\sigma}^{}$ annihilates a conduction electron with moment $k$,
orbital $m$, and spin $\sigma$ in lead $\alpha = L,R$.
Interactions
$U$, 
$W (0 \le W \le U)$, and $J(\le0)$ are the
intra- and interorbital Coulomb repulsion,
and Hund's rule coupling,
respectively.
$n_{dm\sigma}^{}$ is the number of electron in the dot level with orbital $m$
and spin $\sigma$, and ${\bm S}_{dm}^{}$ is
the spin operator 
of electrons in orbital $m$ of the dot.
The intrinsic level width of the dot levels owing to tunnel 
coupling $v_{\alpha}^{}$ 
is given by
$\Gamma = \sum_\alpha \pi \rho_{\rm c}^{} v_\alpha^2$
with
the density of state of the conduction electrons $\rho_{\rm c}^{}$.
The two-orbital case
($M=2$) has been experimentally realized 
in such systems as vertical dots,
carbon-nanotube dots, and four-terminal double-dots
\cite{PhysRevLett.93.017205,NaturePhys5.208,PhysRevLett.101.186804}.
For simplicity, 
the symmetric lead-dot coupling $v_L^{}=v_R^{}$
and the particle-hole symmetry
$\epsilon_{d}^{} = -U/2-(M-1)W$
are assumed.
The chemical potentials of the leads $\mu_{L/R}^{} = \pm V/2$,
satisfying $\mu_L^{} - \mu_R^{} = V(> 0)$,
are measured relative to the Fermi level defined 
at zero voltage $V=0$.
We take units with $\hbar = k_{\rm B} = e =1 $ throughout this Letter.

{\it Full counting statistics}.---
The probability distribution $P(q)$ 
of the transferred charge $q$ across the dot
during a time interval ${\cal T}$ can
provide the current correlation
functions to all orders.
In order to treat them systematically,
we calculate the CGF
$\ln \chi \left(\lambda \right) = \ln \sum_{q} e^{i \lambda q}  P( q)$
in the Keldysh formulation \cite{PhysRevB.70.115305}:
$\ln \chi \left( \lambda \right)
= \ln \left\langle 
 T_{C} S_C^{\lambda}
\right \rangle
$
where
$S_{C}^{\lambda} = T_{C} \exp \left\{ -i \int_{C} dt \left[ {\cal H}_T^{\lambda}(t) + {\cal H}_I^{}(t) \right] \right\}$
is the time evolution operator for an extended Hamiltonian 
${\cal H}_A^{\lambda} = {\cal H}_0^{} + {\cal H}_T^{\lambda}+{\cal H}_I^{}$,
$C$ is the Keldysh contour along
$[t: -{\cal T}/2 \to +{\cal T}/2 \to -{\cal T}/2]$,
$T_{C}$ is the contour-ordering operator,
and $\lambda$ is the counting field.
Here, ${\cal H}_T^{\lambda}$ is given by
\begin{eqnarray}
{\cal H}_T^{\lambda} = \sum_{km\sigma} \left[ v_L^{} e^{i\lambda^{}(t)/2} d_{m\sigma}^{\dagger} c_{kLm\sigma}^{}
+ v_R^{} d_{m\sigma}^{\dagger} c_{kRm\sigma}^{} + \mbox{H.c.} \right] ,
\end{eqnarray}
with the contour-dependent counting-field
defined by
$\lambda (t) \equiv \lambda_{\mp}^{} = \pm \lambda$ 
for the forward and backward 
paths labeled by ``$-$" and ``$+$" respectively.

To calculate the CGF,
we use Komnik and Gogolin's procedure
\cite{PhysRevB.73.195301} outlined below.
First, a more general function
$\chi (\lambda_{-}^{}, \lambda_{+}^{} )$
is introduced.
It is basically given by
$\ln \left\langle 
 T_{C} S_C^{\lambda}
\right \rangle$
but $\lambda_{\mp}^{}$ is formally
treated as an independent variable
assigned for each contour.
For the long time limit ${\cal T} \to \infty$ where 
the switching effect is negligible,
the general CGF is proportional to ${\cal T}$.
Then, the derivative of
the CGF with respect to $\lambda_{-}^{}$ is given
in terms of Green's functions as
\begin{eqnarray}
&& \frac{d}{d\lambda_{-}^{}} \ln \chi (\lambda_{-}^{},\lambda_{+}^{}) \nonumber \\
&& \quad = -i {\cal T} \frac{v_L^2}{2} \sum_{km\sigma} \int \frac{d\omega}{2\pi}
\left[ e^{-i \bar{\lambda}/2} G_{d}^{\lambda -+} \left( \omega \right) g_{kL}^{0+-} \left( \omega \right) \right. \nonumber \\
&&\quad \qquad \qquad \qquad \qquad - \left. e^{i \bar{\lambda}/2} g_{kL}^{0-+} \left( \omega \right) G_{d}^{\lambda +-} \left( \omega \right) \right] , \label{eq:dAP2}
\end{eqnarray}
with $\bar{\lambda} \equiv \lambda_{-}^{} - \lambda_{+}^{}$.
$g_{k\alpha}^{0-+}(\omega)= i2\pi \delta(\omega -\varepsilon_{k\alpha}^{}) f_{\alpha}^{}(\omega)$ and
$g_{k\alpha}^{0+-}(\omega)= -i2\pi \delta(\omega -\varepsilon_{k\alpha}^{}) [1 - f_{\alpha}^{}(\omega)]$ are
the lesser and greater parts of the Green's function for
electrons in lead $\alpha$, respectively,
with the Fermi distribution function
$f_{\alpha}^{} (\omega) = [e^{(\omega - \mu_{\alpha}^{})/T} + 1]^{-1}$.
For the long time limit ${\cal T} \to \infty$,
the dot Green's function is defined as
$G_{d}^{\lambda \, \nu\nu'} (\omega)
= -i\int d(t-t') e^{i\omega (t-t')} \langle T_{C}^{} 
d_{m\sigma}^{}(t_{\nu}) d_{m\sigma}^{\dagger} (t'_{\nu'}) 
\rangle_{\lambda_-,\lambda_+}^{}$.
$\nu$ and $\nu'$ are the labels for the two Keldysh contours.
Here, we use a notation
$
\langle A(t) \rangle_{\lambda_{-}^{},\lambda_{+}^{}}^{}
\equiv
\left\langle 
T_{C}^{}
S_C^{\lambda} A(t)
\right\rangle 
\big/ \chi ( \lambda_{-}^{}, \lambda_{+}^{})
\label{eq:exexpectation}
$
which represents an expectation for Hamiltonian
${\cal H}_A^{\lambda}$.
Once $\ln \chi (\lambda_{-}^{}, \lambda_{+}^{})$ is computed,
the statistics are recovered from
$\ln \chi ( \lambda )
= \ln \chi ( \lambda,- \lambda )$.

{\it Renormalized perturbation theory}.---
To calculate the Keldysh Green's function  
with finite $\lambda$ at low energies, 
we use the RPT outlined below
\cite{PhysRevLett.70.4007,PhysRevB.82.115123,JPSJ.74.110}.
The basic parameters that specify the impurity Anderson model ${\cal H}_A^{}$
are $\epsilon_{d}$, $\Gamma$, $U$, $W$, and $J$.
Correspondingly, the low-energy
excitations can be characterized by
the renormalized parameters  for the quasiparticles: dot-level 
$\widetilde{\epsilon}_{d}^{} = z \left[ \epsilon_{d}^{} + \Sigma_{d}^r (0) \right]$,
level width
$\widetilde{\Gamma} = z \Gamma$, where  $\Sigma_{d}^r (\omega)$ is the self-energy of the retarded Green's function
for the dot state 
$G_{d}^r (\omega) = [ \omega - \epsilon_{d}^{} + i \Gamma - \Sigma_{d}^r (\omega) ]^{-1}$,
$z = [ 1 - \partial \Sigma_{d}^{r}(\omega)/\partial \omega|_{\omega=0} ]^{-1}$
is the wave function renormalization factor.
Here, $\widetilde{\Gamma}$ corresponds to
the characteristic energy scale, namely,
the Kondo temperature $T_K=\pi \widetilde{\Gamma}/4$.
The quasiparticle interactions
$\tilde{U}, \tilde{W}$ and $\tilde{J}$ are defined in terms of the local full four-vertex for the scattering of the electrons,
$\Gamma_{m_3 \sigma_3; m_4 \sigma_4}^{m_1 \sigma_1; m_2 \sigma_2} (\omega_1, \omega_2, \omega_3, \omega_4)$
taken at zero frequency $\omega_i=0$
(see Ref.\ \onlinecite{PhysRevB.82.115123}).
In this Letter,
we choose the 
renormalized parameters defined
at the equilibrium ground state.
Then, the replacement of the bare parameters with the
renormalized ones gives the leading terms of an effective
Hamiltonian corresponding to the low-energy fixed point
of the numerical renormalization group (NRG).
The renormalized  parameters 
can be
related to the  enhancement
factors of spin, orbital and charge susceptibilities;
$z \widetilde{\chi}_{\rm s}^{} =
1+ \widetilde{\rho}_{d}^{}(0) \left[ \widetilde{U}-(M-1)\widetilde{J} \right],
z \widetilde{\chi}_{\rm orb}^{} =
1+ \widetilde{\rho}_{d}^{}(0) \left[ 2\widetilde{W}-\widetilde{U} \right], 
z \widetilde{\chi}_{\rm c}^{} =
1- \widetilde{\rho}_{d}^{}(0) \left[ \widetilde{U}+2(M-1)\widetilde{W} \right] ,$
with the renormalized density of state
$\widetilde{\rho}_{d}(\omega)= (\widetilde{\Gamma}/\pi)/[(\omega - \widetilde{\epsilon}_{d}^{})^2+ \widetilde{\Gamma}^2]$, 
and also to 
the Friedel sum rule 
$\pi \langle n_{dm\sigma}^{}\rangle = \cot^{-1} \left( \widetilde{\epsilon}_{d}^{}/ \widetilde{\Gamma} \right)$.
These relations enable one to deduce the value
of renormalized parameters in some special limits.
For instance, the renormalized dot level in the particle-hole symmetric case
is situated at the Fermi energy $\tilde{\epsilon}_d=0$.
We can also evaluate
the renormalized parameters for dots with two orbitals
$(M=2)$
with the NRG approach
\cite{PhysRevB.82.115123}.

The nonequilibrium perturbation theory in powers of $U,W$ and $J$
can be reorganized as 
an expansion with respect to the renormalized interactions
$\widetilde{U}, \widetilde{W}$, and $\widetilde{J}$
by taking the free quasiparticle 
Green's function of the form
\begin{eqnarray}
\widetilde{g}_{d}^{\lambda --} (\omega) &=&
\left[ \omega - \widetilde{\epsilon}_d^{} - i\widetilde{\Gamma}+ i\widetilde{\Gamma} \textstyle\sum_l f_l^{} \right] \Big/ {\cal D} , \nonumber \\ 
\widetilde{g}_{d}^{\lambda -+} (\omega) &=&
\left[ i \widetilde{\Gamma} e^{i \bar{\lambda}/2} f_L^{}
+ i \widetilde{\Gamma} f_R^{} \right] \Big/{\cal D} , 
\nonumber \\
\widetilde{g}_{d}^{\lambda +-} (\omega) &=&
- \left[ i \widetilde{\Gamma} e^{-i \bar{\lambda}/2} \left( 1-f_L^{} \right)
+ i \widetilde{\Gamma} \left( 1-f_R^{} \right) \right] \Big/ {\cal D} ,  
\nonumber \\
\widetilde{g}_{d}^{\lambda ++} (\omega) &=&
\left[ - \omega + \widetilde{\epsilon}_d^{} - i\widetilde{\Gamma} + i\widetilde{\Gamma} \textstyle\sum_l f_l^{}  \right] \Big/ {\cal D},
\label{eq:zeropart}
\end{eqnarray}
with
${\cal D} = \left( \omega -\widetilde{\epsilon}_d^{} \right)^2 + \widetilde{\Gamma}^2
+ \widetilde{\Gamma}^2  \left( e^{i \bar{\lambda}/2} - 1 \right) 
\left( 1 - f_R^{} \right) f_L^{}$
as the zero-order propagator
\cite{JPSJ.74.110}.
With this approach,
the exact form of 
the Green's function can be calculated
at low energies up to terms of order 
$\omega^2$, $V^2$, and $T^2$.

We calculate the 
Green's function ${\bm G}_{d}^{\lambda}(\omega)$
under a finite counting field through
the Dyson equation
\begin{eqnarray}
{\bm G}_{d}^{\lambda} (\omega)
= z \widetilde{{\bm G}}_{d}^{\lambda} (\omega)
= z \left[ { \widetilde{{\bm g}}_{d}^{\lambda} (\omega)}^{-1}
- \widetilde{{\bm \Sigma}}_{d}^{\lambda} (\omega) \right]^{-1}.
\label{eq:EXGF}
\end{eqnarray}
The renormalized self-energy at $T=0$ up to $\omega^2, \omega V$, and $V^2$
in the particle-hole symmetric case
can be calculated in the second order perturbation
in the three renormalized interactions as
\begin{eqnarray}
\widetilde{{\bm \Sigma}}_{d}^{\lambda} (\omega) =
\frac{-i}{8 \widetilde{\Gamma}} \, \widetilde{\cal I}
\left[
\begin{array}{cc}
A(\omega, V) & B(\omega, V) \\
- B_{}^{\ast}(-\omega, -V) & A (\omega, V)
\end{array}
\right],
\label{eq:EXSE}
\end{eqnarray}
with
\begin{eqnarray}
A (\omega,V) &=&
\left[ a\left(\omega, 3V/2 \right) + 3\, a\left( \omega, V/2 \right) \right.  \nonumber \\
 && \qquad+ \left. 3\, a\left( \omega, -V/2 \right) + a\left( \omega,-3V/2 \right) \right], \\
B(\omega, V) &=&
\left[ e^{-i \bar{\lambda}/2} b\left( \omega,3V/2 \right) 
+ 3 b \left( \omega, V/2 \right) \right. \nonumber \\
&& \  \left. + 3 e^{i \bar{\lambda}/2}   b\left( \omega, - V/2 \right)
+  e^{i \bar{\lambda}} b \left( \omega,-3V/2 \right) \right], \\
\widetilde{\cal I} &=&
 \tilde{u}^2
+ 2(M-1) \left( \tilde{w}^2
+ 3 \tilde{j}^2 /4 \right)
\label{eq:IPofSigma}
\end{eqnarray}
Here,
$a(\omega,x)=- \frac{1}{2}(\omega - x)^2 \mbox{sgn}(-\omega + x)$,
$b(\omega,x)=(\omega -x)^2 \theta(-\omega +x)$,
and $\tilde{u} \equiv \widetilde{U}/(\pi\widetilde{\Gamma}),
\tilde{w} \equiv \widetilde{W}/(\pi\widetilde{\Gamma})$
and $\tilde{j} \equiv \widetilde{J}/(\pi\widetilde{\Gamma})$.
For $\lambda = 0$, 
it corresponds to an extension 
of Ref.\ \onlinecite{PhysRevB.64.153305}
to the orbital-degenerate Anderson model.

{\it Results and discussion}.---
Integrating Eq.\ (\ref{eq:dAP2}) with respect to
$\lambda_{-}^{}$,
the CGF is calculated at $T=0$ 
up to $V^3$, as $\ln \chi (\lambda)
= {\cal F}_0^{} + {\cal F}_1^{} + {\cal F}_2^{}$
with
\begin{eqnarray}
{\cal F}_0
&=& \frac{2M {\cal T}}{2\pi} \int_{-V/2}^{V/2} d\omega
\ln\left[ 1 + T(\omega) \left( e^{i \lambda } -1 \right) \right] ,
\label{eq:F0} \\
{\cal F}_1
&=& {\cal T} P_{b1}
\left( e^{-i\lambda} -1 \right), \quad
{\cal F}_2
= {\cal T} P_{b2}
\left( e^{-i2 \lambda } -1 \right) .
\end{eqnarray}
Here, ${\cal F}_0^{}$
describes
free-quasiparticle tunneling via the renormalized level
with transmission probability
$T(\omega)=\widetilde{\Gamma}^2/(\omega^2 + \widetilde{\Gamma}^2)$,
which carries single charge $e$.
${\cal F}_1^{}$ and ${\cal F}_2^{}$
represent the scattering process 
due to the residual interactions
$\widetilde{U}$, $\widetilde{W}$, and $\widetilde{J}$.
From the coefficient of the counting field in
${\cal F}_1^{}$ and ${\cal F}_2^{}$,
it is clear that
$P_{b1} =\frac{M}{12\pi} \tilde{\cal I} \frac{V^3}{\widetilde{\Gamma}^2}$
and
$P_{b2} = 
\frac{M}{6\pi} \tilde{\cal I} \frac{V^3}{\widetilde{\Gamma}^2}$
are the probabilities of the single- and paired-quasiparticle 
backscattering processes carrying charge $e$ and $2e$,
respectively.
Then, the $n$th-order 
cumulant ${\cal C}_n^{}= (-i)^n\frac{d_{}^n}{d\lambda_{}^n}\ln \chi(\lambda)$
is readily derived as
\begin{eqnarray}
{\cal C}_n^{}= {\cal T} \left[ I_{u} \delta_{1n} + (-1)^n  (P_{b0} + P_{b1} + 2^{n} P_{b2}) \right], \label{eq:FullCCumulant}
\end{eqnarray}
where $I_u = 2MV/(2\pi)$ is the linear-response current
and $P_{b0} =\frac{M}{12\pi} \frac{V^3}{\widetilde{\Gamma}^2}$
represents the probability of the single-quasiparticle 
backscattering processes due to the renormalized level.
Especially, the factor $2^n$ characterizing cumulant
of interacting electrons (\ref{eq:FullCCumulant})
indicates the existence of 
spin-singlet, orbital-singlet, 
and  spin-triplet pairs 
of quasiparticles in the nonequilibrium current.
Equation (\ref{eq:FullCCumulant}) generalizes 
the previous results of
the cumulant for the SU($2M$) case where $U=W$ and $J=0$
\cite{PhysRevLett.97.016602,PhysRevB.76.241307,PhysRevB.83.241301}.
In the present result, the Hund's coupling
enters through the $P_{b1}$ and $P_{b2}$ 
as well as the Kondo energy scale $\tilde{\Gamma}$. 
Therefore, 
it gives rise to  
the spin-triplet pairs in the current,
and also varies the couplings  
of the  spin sector $\tilde{u}$ 
and the orbital sector $\tilde{w}$, 
which causes the crossover seen in the current fluctuation 
as discussed below.

In the series of cumulant ${\cal C}_n^{}$, 
the first two coefficients 
${\cal C}_1^{}$  and ${\cal C}_2^{}$ give 
time-averaged current $I = {\cal C}_1^{}/{\cal T}$ 
and shot noise $S=2{\cal C}_2^{}/{\cal T}$, respectively.
The higher order ones, ${\cal C}_n^{}$ for $n \geq 3$, are determined by 
these two
and are automatically generated by Eq.\ (\ref{eq:FullCCumulant}).
From these coefficients, 
the Fano factor of the shot noise for the backscattering current
$I_b \equiv I_u - I$, can be deduced in the form
\begin{eqnarray}
F_b \,\equiv \, 
\frac{S}{2I_b}
\,=\, \frac{1+9\,\widetilde{{\cal I}}}{1 + 5\,\widetilde{{\cal I}}}.
\label{eq:FF}
\end{eqnarray}
This is an important result of this Letter
\footnote{
Our calculations can be extended to the particle-hole {\it asymmetric} case
but the equations are  more complicated as  the transport coefficients
depend also on the order $\omega^2$ term of the
{\it real part} of the self-energy
\cite{PhysRevB.64.153305}.}.
The Fano factor $F_b$ determines
the effective charges 
of the backscattering current
\cite{PhysRevLett.97.086601}.
Our results, Eq.\  (\ref{eq:FF}) with (\ref{eq:IPofSigma}), 
show that $F_b$ also represents the degree of 
the residual interaction,
as the Wilson ratio $R = z \widetilde{\chi}_{\rm s}$ does in the SU(2) case.
We examine the value of the Fano factor in some special limits 
in the following.

First, in the weak coupling limit
$u,w, |j| \ll 1$,
the renormalized parameters
take a form $(\widetilde{u}, \widetilde{w}, \widetilde{j}) \to (u,w,j)$ and $z \to 1- \left( 3-\frac{1}{4}\pi^2 \right) \left[ u^2 + 2(M-1) \left( w^2 + 3j^2/4 \right) \right]$
with scaled interactions
$u \equiv U/(\pi\Gamma),w \equiv W/(\pi\Gamma)$, and $j \equiv J/(\pi\Gamma)$.
Substituting these 
into Eq.\ (\ref{eq:FF}), the result for 
the second-order perturbation in the bare interactions is produced.
In particular, the Fano factor for noninteracting limit $u=w=j=0$ naturally becomes unity which represents single charge transport 
given by Eq.\ (\ref{eq:F0}).

In the large Hund's coupling limit $|j| \gg 1$,
the dot state is restricted to the highest spin state, 
for which we obtain the Kondo coupling ${\cal H}_K^{}$ from 
the Anderson model ${\cal H}_A$ in the form
\begin{eqnarray}
{\cal H}_K^{} =J_{K}^{} 
\sum_{kk'\sigma\sigma'm} c_{km\sigma}^{\dagger} 
{\bm \sigma}_{\sigma\sigma'} c_{k'm\sigma'}^{} \cdot{\bm S}_{d}
\label{eq:KondoH}
\end{eqnarray}
with $J_{K}^{} =4(v_L^2 + v_R^2)\left[U-(M-1)J \right]^{-1}$,
the Pauli matrix ${\bm \sigma}_{\sigma\sigma'}$,
$c_{km\sigma}^{} =
\sum_{\alpha} c_{k\alpha m\sigma}^{} /\sqrt{2}$,
and the $S=M/2$ spin operator for the dot-site ${\bm S}_d$.
In this limit,
the high spin $S=M/2$ $(\ge 1)$ state is fully screened,
and the Wilson ratio becomes $R =2(M+2)/3$ \cite{PTP.55.67}.
The charge and orbital fluctuations 
are suppressed $\widetilde{\chi}_{\rm c} \to 0$ and $\widetilde{\chi}_{\rm orb} \to 0$ 
in the limit $j \to -\infty$. 
Therefore, the renormalized parameters take 
the value $(\widetilde{u}, \widetilde{w}, \widetilde{j}) \to (1, 0, - 2/3)$.
Then,  through Eqs.\ (\ref{eq:IPofSigma}) 
and  (\ref{eq:FF}), we find the Fano factor
in this limit,  
\begin{eqnarray}
F_b
\to \frac{9M+6}{5M+4}= \frac{9S+3}{5S+2}.
\label{eq:FFstrongJ}
\end{eqnarray}
We note that to observe 
the Fano factor due to the higher-spin Kondo effect 
in real experiments,
Hund's coupling $J$ is not required to be very large 
as long as it is larger than the Kondo temperature $J > T_K$.
%
%
The explicit values of the Fano factor in this limit
for several $M$ are shown in Table \ref{tab:uRatio}.
For comparison,
the Fano factor in the SU($2M$) Kondo limit $u=w \to \infty, j=0$,
$F_b \to (M + 4) / (M + 2)$
\cite{PhysRevB.80.155322,PhysRevB.83.241301},
is also shown in Table \ref{tab:uRatio}.
\begin{table}[tb]
\caption{\label{tab:uRatio}
The Fano factor $F_b={\cal C}_2^{}/({\cal T}I_b^{})$
in
(a) the large Hund's coupling
$j\to -\infty$ or the $S=M/2$ Kondo limit,
and (b) the SU($2M$) Kondo limit
$u=w\to\infty, j=0$.
}
\begin{ruledtabular}
\begin{tabular}{lccccc}
$M$ & 1 & 2 & 3 & 4 & $\to \infty$ \\
\hline
(a) $S=M/2$ Kondo
&  & 12/7 & 33/19 & 7/4 & $\to 9/5$ \\
(b) SU($2M$) Kondo & 5/3 & 3/2 & 7/5 & 4/3 & $\to 1$
\end{tabular}
\end{ruledtabular}
\end{table}
Increasing orbital degeneracy $M$ in the SU($2M$) Kondo limit,
the system approaches the mean-field limit $F_b \to 1$
\cite{PhysRevB.83.241301}.
In contrast, for large $|j|$,
the renormalized interactions
$\tilde{u}$, $\tilde{w}$, and $\tilde{j}$
converge to values independent of the
orbital degeneracy $M$.
As $M$ increases, the number of coupled orbitals
with the Hund's coupling $J$ increases.
It results in an enhancement of the spin fluctuations, 
as seen in the expression of
$z\widetilde{\chi}_{\rm s}$
or the interaction factor of the self-energy given 
in Eq.\ (\ref{eq:IPofSigma}).
Therefore, we conclude that 
the Fano factor increases with $M$, 
through the enhancement of the quasiparticle-pair scatterings 
due to the residual Hund's coupling $\tilde{j}$.

\begin{figure}[tb]
\includegraphics[width=7.6cm]{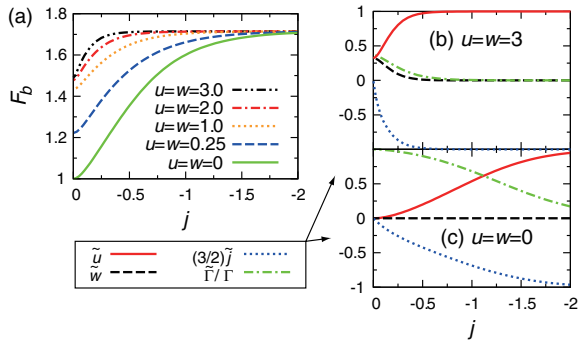}
\caption{
\label{fig:FF_U=W}
(Color online)
(a) The Fano factor for several choices of Coulomb repulsion $u=w$
as a function of scaled Hund's coupling $j$.
(b) The renormalized parameters for $u=w=3$ and (c) $u=w=0$,
as a function of $j$.
}
\end{figure}

\begin{figure}[tb]
\includegraphics[width=7.6cm]{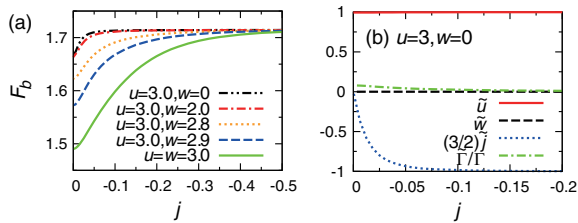}
\caption{
\label{fig:FF_U=3_W=C}
(Color online)
(a) The Fano factor for several Coulomb interaction $w (\le u=3)$
as a function of scaled Hund's coupling $j$.
(b) The renormalized parameters for $(u,w)=(3,0)$
as a function of $j$.
}
\end{figure}

\begin{figure}[tb]
\includegraphics[width=7.6cm]{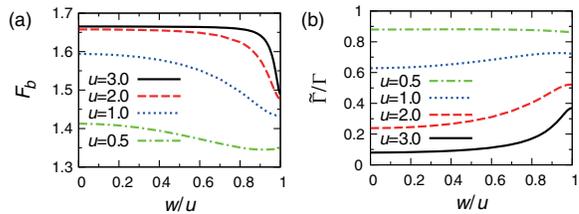}
\caption{
\label{fig:FF_J=0_U=C}
(Color online)
(a)The Fano factor and
(b) the renormalized level width for no Hund's coupling $j=0$
and several intraorbital Coulomb repulsions $u$,
as a function of interorbital Coulomb repulsions $w (\le u)$.
}
\end{figure}

Next, $J$-dependent crossover of the Fano factor
for the two-orbital case ($M=2$) is investigated,
evaluating the renormalized parameters with the NRG approach.
The Fano factor for $u=w$ and $u=3>w$
as a function of Hund's coupling $j$ are shown
in Fig.\ \ref{fig:FF_U=W}(a)
and Fig.\ \ref{fig:FF_U=3_W=C}(a), respectively.
The crossover is observed in the Fano factor 
around the coupling corresponding to
the Kondo temperature 
$|j| \sim t_K \equiv \widetilde{\Gamma}/(4\Gamma)$.
At large $|j|$, the Fano factor for any $u\ge w$ converges
to a value 12/7 given by Eq.\ (\ref{eq:FFstrongJ}),
as discussed above.
However, at small $|j|$, the values of the Fano factor depend
on Coulomb repulsions $u$ and $w$.
The Fano factor and the renormalized level-width (or the Kondo temperature)
at $j=0$ for several intraorbital Coulomb interaction $u$
as a function of interorbital
Coulomb interaction  $w$ are shown in Fig.\ \ref{fig:FF_J=0_U=C} (a) and (b), respectively.
With an increase of interorbital Coulomb repulsion $w$,
the Fano factor varies 
from the SU(2) value to the SU(4) value
around $(u-w) \sim t_K$.

The crossover seen in
the Fano factor is a consequence of the behavior of the 
renormalized parameters, which enter through Eq.\ (\ref{eq:FF}).
Thus, the feature discussed above is also seen in
the renormalized parameters,
which for $(u,w)=(3,3)$, $(0,0)$, and $(3,0)$
are shown in Figs.\ \ref{fig:FF_U=W} (b) and (c),
and Fig.\ \ref{fig:FF_U=3_W=C} (b), respectively,
as a function of $j$.
Particularly, in Fig.\ \ref{fig:FF_U=W} (b),
we see a general trend that the spin coupling of
the Kondo correlation $\widetilde{u}$ increases,   
whereas the orbital coupling $\widetilde{w}$ decreases,
with increase of the ferromagnetic coupling $-j$.
These renormalized parameters converge
to the value $(\tilde{u}, \tilde{w}, \tilde{j})=(1,0,-2/3)$ 
in the large Hund's coupling limit.
Even for $(u,w)=(0,0)$, the Hund's coupling enhances
the spin coupling $\tilde{u}$ 
as shown in Fig.\ \ref{fig:FF_U=W}(c).

{\it Summary}.---
We have studied the role of 
the ferromagnetic Hund's rule coupling 
on the FCS 
for the orbital-degenerate Anderson impurity
in the particle-hole symmetric case.
Using the RPT,
we have derived the CGF for nonequilibrium-current distribution. 
The result is asymptotically exact at low energies and 
is described by the quasiparticles of the local Fermi liquid.
Specifically, the explicit expression of the Fano factor for 
the shot noise is obtained
in the
fully screened higher-spin Kondo limit,
which depends only on the orbital degeneracy $M$.
We have also investigated the crossover between the 
large and small Hund's coupling limits
in the two-orbital case ($M=2$), using the NRG approach.
Furthermore, the CGF indicates that 
the Hund's coupling gives rise to singlet and triplet pairs of quasiparticles
carrying charge $2e$ in the backscattering current
and these correlated charges characterize
the current fluctuation through the dot.

R.S. acknowledges
Takeo Kato, Yasuhiro Utsumi, Kensuke Kobayashi,
Yuma Okazaki, and Satoshi Sasaki
for fruitful discussion.
This work was supported by
the JSPS through its FIRST program,
the JSPS Grant-in-Aid
for Scientific Research C (No.\ 23540375) and S (No.\ 19104007),
and MEXT through KAKENHI ``Quantum Cybernetics" project and through Project for
Developing Innovation Systems.


%

\end{document}